\documentclass[prl, aps, showpacs,tightenlines, twocolumn, amsmath,amssymb]{revtex4}
\usepackage[dvips]{graphicx,color}
\usepackage{dcolumn}
\usepackage{bm}

\begin{document}

\preprint{APT/123-QED}

\title{Amorphous state in the mixed phase of hadron-quark phase transition
\\ in protoneutron stars}%

\author{Nobutoshi Yasutake$^{1}$}
 \email{nobutoshi.yasutake@it-chiba.ac.jp}
\author{Toshiki Maruyama$^2$}
\author{Toshitaka Tatsumi$^3$} 
\affiliation{%
$^1$Physics Department, Chiba Institute of Technology, Shibazono 2-1-1, Narashino, Chiba 275-0023, Japan\\
$^2$Advanced Science Research Center, Japan Atomic Energy Agency, Tokai, Ibaraki 319-1195, Japan\\
$^3$Department of Physics, Kyoto University, Kyoto 606-8502, Japan
}%

\date{\today}

\begin{abstract}
We study the hadron-quark mixed phase in protoneutron stars, where neutrinos are trapped and lepton number becomes a conserved quantity besides the baryon number and electric charge. Considering protoneutron-star matter as a ternary system, the Gibbs conditions are applied together with the Coulomb interaction. 
We find that there no crystalline (``pasta") structure appears  in the regime of high lepton-number fraction; the size of pasta becomes very large and the geometrical structure becomes mechanically unstable due to the charge screening effect. 
Consequently the whole system is separated into two bulk regions like an amorphous state, where the surface effect is safely neglected. There, the local charge neutrality is approximately attained, so that the equation of state is effectively reduced to the one for a binary system. Hence, we conclude that there is no 
possibility for the density discontinuity 
to appear in protoneutron-star matter, which is a specific feature in a pure system. 
These features are important when considering astrophysical phenomena such as supernova explosions or radiation of the gravitational wave from protoneutron stars.
\end{abstract}

\maketitle

Recently there have been many works about the hadron-quark~(HQ) phase transition in compact stars \cite{haensel07}, where signatures of the phase transition are explored 
during supernova explosions~\cite{sagert09,ficher11}, in the spectrum of the gravitational wave \cite{sotani11} and so on.
The deconfinement transition might be of the first order at finite density although  uncertainties are left, 
e.g., the equation of state (EOS) of quark matter or deconfinement mechanism \cite{fukushima11}. 
In this case, the HQ-mixed phase appears and
the phase equilibrium in the mixed phase must be carefully treated 
by applying the Gibbs conditions \cite{glendenning92}. 
A simple treatment of the mixed phase may be the bulk Gibbs calculation, 
where phase equilibrium of two pieces of bulk matter is considered without the electromagnetic interaction. 
The EOS then looks like the one of the second-order phase transition in the sense of Ehrenfest; 
there appears no constant-pressure region as in the van der Waals liquid. 
A realistic treatment of the mixed phase takes into account the ``{\it finite-size effects}",   the electromagnetic interaction and the surface tension.
Generally the properties of the mixed phase strongly depend on the finite-size effects to lead to the inhomogeneous crystalline (pasta) structures \cite{ravenhall83,heiselberg93}. 
The EOS resembles the one given by the bulk Gibbs calculation for the weak surface tension, and close to the one given by the Maxwell construction for the strong surface tension~\cite{endo06, maruyama07}. 
These differences are mainly brought about by the charge-screening effect and rearrangement of the charge distribution in the presence of the Coulomb interaction\cite{voskresensky02}. 

Here we consider the HQ-phase transition in protoneutron stars(PNSs), where neutrinos are trapped and lepton number is conserved. 
Consequently there are three conserved quantities, ---baryon number, electric charge, and lepton number. 
Thus, PNS matter should be treated as a ternary system, 
different from the usual neutron star matter which is a binary system. 
The Gibbs conditions can also be applied to this case as a straight extension from the previous works.
In the recent papers, a new type of the mixed phase has been suggested on the HQ-phase transition in PNSs~\cite{pagliara09, hempel09}. 
They derived an EOS with the HQ mixed phase by treating PNS matter as a binary system, so that there appears no density discontinuity  though the phase transition. It is an interesting scenario and gives rise to an important ingredient in the numerical simulation of supernova explosions\cite{sagert09,ficher11}. However, there remains a serious uncertainty which comes from the finite-size effects, since they have simply carried out a bulk Gibbs calculation without any finite-size effect by imposing the local charge neutrality condition {\it a priori}. Hence, it is very important to examine them carefully.
In this paper, we study the HQ mixed phase with the finite-size effects by extending the previous works~\cite{maruyama07,yasutake09b}. 
Then, we compare our results to other results using the bulk calculations with the local charge neutrality~(LCN) and the global charge neutrality~(GCN) conditions.

The theoretical framework for the hadronic phase of matter is provided by the
Brueckner-Hartree-Fock approach including hyperons such as
$\Sigma^-$ and $\Lambda$~\cite{burgio11}.
For the inclusion of the thermal effects, we adopt the frozen-correlation approximation at finite temperature~\cite{yasutake09b,burgio11,baldo99}. 
This approximation has been proved to be feasible for temperature $T < 50$ MeV.

For the quark phase, we adopt the density-dependent bag model consisting of
$u,d,s$ quarks using the same parameters as Ref.~\cite{nicotra06}. 
This model is probably too simple to describe quark matter in a realistic way; 
we will adopt more sophisticated models in the future~\cite{yasutake09a}.
We assume massless $u$ and $d$ quarks, and $s$ quarks with the current mass of $m_s=150$ MeV. 

Electrons and neutrinos are present in both phases.
For simplicity, muons and antiparticles are not included in this paper.

Then all the chemical potentials can be written in terms of three independent chemical potentials---the baryon number ($\mu_B$), the charge ($\mu_C$), and the lepton number ($\mu_L$) ones
\begin{eqnarray}
&& \mu_n= \mu_\Lambda = \mu_B ,~~ 
\mu_p=\mu_B+\mu_C, \nonumber\\
 && \mu_{\Sigma^-} + \mu_p = 2\mu_B ,
\label{chemb}
\end{eqnarray} 
for baryons, and
\begin{eqnarray}
\label{chemql}
 && \mu_u= \frac{1}{3}\mu_B+\frac{2}{3}\mu_C,~~
 \mu_d=\mu_s=\frac{1}{3}\mu_B-\frac{1}{3}\mu_C, \nonumber\\
 && \mu_{\nu_e}=\mu_L, ~~\mu_e=\mu_L-\mu_C ,
\end{eqnarray}
for quarks and leptons.
Thus, we can see PNS matter is a ternary system. We can easily extend our previous procedure to study the phase equilibrium in the mixed phase, 
which has been successfully applied to a binary system \cite{yasutake09b}.
Using the local density approximation for particles, we have the thermodynamic potential and the coupled equations for particle densities and the Coulomb potential $V_C({\bf r})$. 
Introducing the Wigner-Seitz cell appropriate to various geometrical structures (pasta)---droplet, rod, slab, tube, and bubble, 
we solve the coupled equations self-consistently under the Gibbs conditions. 
For the interface between the hadron and quark phases we put a sharp boundary with a constant surface tension parameter, $\sigma_S$. 
We present our results by using a typical value of $\sigma_S=$40 MeV~fm$^{-2}$, 
discussing its dependence later. 

We also carry out two kinds of calculation for comparison: one is the bulk Gibbs calculation  with a GCN without the finite-size effects for a  ternary system, 
which assumes two pieces of bulk matter endowed with different baryon-number density, charge and lepton-number fraction ($Y_l$). The other is also a bulk calculation, but LCN  is assumed. 
Thereby, we violate the Gibbs condition for charge chemical potential, in other words, we have different values of electron chemical potential in both phases, $\mu_e^H\neq\mu_e^Q$. Note that the matter in this case becomes essentially a binary system; 
it is further reduced to a pure system in the absence of neutrinos ($\mu_{\nu_e}=0$),  and there appears a constant pressure region in the EOS.

For the PNS matter, we must consider mainly two effects, namely temperature and neutrino fraction. The range of temperature is roughly $0 < T< 30$ MeV, and the range of lepton-number fraction is $0.05 < Y_l < 0.4$~\cite{burrows86}. At the final stage of the evolution of PNSs, the neutrino fraction $Y_\nu$ becomes almost zero. 
 Hence, we calculate two extreme cases in this paper, PNS matter of $T=30$ MeV and $Y_l = 0.4$.

 First, we show the pressure of  PNS matter as a function of baryon-number density ($n_B$) in 
Fig.~\ref{fig:01}.
As we shall see below, there is no minimum in the free energy as a function of the cell-size $R_W$, i.e., $R_W\rightarrow \infty$, which means the crystalline structures of pasta are broken there to form  ``an amorphous state'' composed of two species of matter; the amorphous state should take a bicontinuous structure with a complex pattern, depending on the kinetics of its formation.
For a practical reason, we fix the cell-size $R_W$ as a large but finite value of $R_W=$
100 fm in the evaluation of the EOS. With this procedure we can clearly see how the amorphous state appears in the large cell-size limit of the pasta structures.
The EOS is then close to the one given by the bulk calculation (LCN). 
This is because the Coulomb energy and the surface tension do not contribute to the total free energy due to the large cell-size $R_W$.
Thus the charge chemical potential becomes irrelevant in this case and 
PNS matter is reduced to a binary system specified by the lepton and baryon-number chemical potentials; actually the EOS looks like the one given by the bulk Gibbs calculation for a binary system.
%
\begin{figure}[h]
\includegraphics[width=18pc]{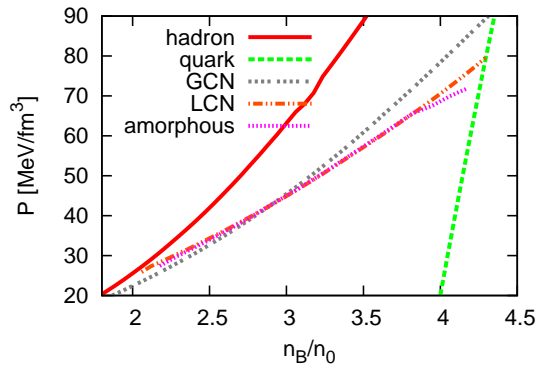}
\caption{\label{fig:01} (color on line).  Pressure vs. density of the amorphous state compared with the bulk Gibbs calculations (GCN,LCN)  for PNS matter at $T=30$MeV and for $Y_l=0.4$. 
The nuclear saturation density is denoted by $n_0$ ($n_0 = 0.17$ fm$^{-3}$).
}
\end{figure}

Readers may be confused at this result since the treatments of the electron chemical potential in the quark and hadron phases are different. Note that the Gibbs condition $\mu_e^H=\mu_e^Q=\mu_e$ never means the equality of the electron number density $n_e$ between both phases. 
\begin{figure}[h]
\includegraphics[width=21pc]{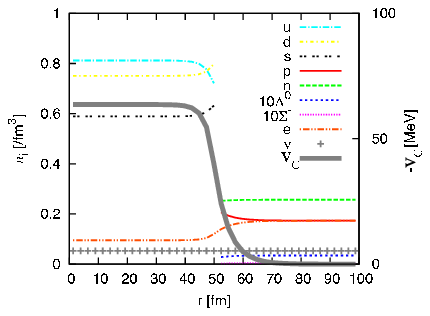}
\caption{\label{fig:02} 
(color on line). Each particle density $n_i$ and the Coulomb potential $V_C$ in the amorphous sate at $n_B = 3 n_0$. We assume a rod-type structure with $R_W=100$ fm in this calculation. 
The density profiles are shown as $u$, $d$, $s$ for $u$, $d$, $s$-quarks,  $p$, $n$, $\Lambda^0$, $\Sigma^-$ for protons, neutrons, $\Lambda^0$ and $\Sigma^-$ hyperons, and $e$, $\nu$ for electrons and electron-type neutrinos. 
}
\end{figure}
By solving Poisson's equation, we properly consider the Coulomb potential consistent  with the density profiles of the particles.
Then $n_e$ must be represented in the gauge invariant form, $n_e=2\int_0^\infty d^3p/(2\pi)^3f_e({\bf r},{\bf p};\mu_e)$ with the Fermi-Dirac distribution function, 
$f_e({\bf r},{\bf p};\mu_e)=[1+{\rm exp}((p-(\mu_e-V_C({\bf r})))/T)]^{-1}$ in the massless limit.
Although the Coulomb interaction is irrelevant for the case, $R_W\rightarrow \infty$, it never means $V_C=0$ in both phases; $V_C$ can take different constants $C^{H,Q}$, 
instead utilizing the gauge degree of freedom \cite{voskresensky02}.
Hence, $n_e^H\neq n_e^Q$ even if $\mu_e^H=\mu_e^Q=\mu_e$, which is realized in this case. 
Furthermore, $\mu_e^H$ and $\mu_e^Q$ in the bulk calculation (LCN) 
correspond to $\mu_e-C^H$ and $\mu_e-C^Q$, respectively.
Actually, the shape of the Coulomb potential almost looks like the Heaviside function. 
Figure~\ref{fig:02} clearly shows this situation, where the density profile of each particle and the Coulomb potential are drawn 
in the amorphous state at $n_B = 3 n_0$.
%
Consequently, local charge neutrality is almost attained in the amorphous state and our EOS qualitatively agrees with 
the one in Refs.~\cite{pagliara09, hempel09}, even if we treat PNS matter as a ternary system. 
Note that lepton-number fraction as well as baryon-number density is different for two phases in the amorphous state.
Hence, we conclude that there is no possibility for the density discontinuity  to appear in EOS which is a specific feature within the Maxwell construction, considering the finite-size effects for PNS matter.

\begin{figure}[h]
\includegraphics[width=18pc]{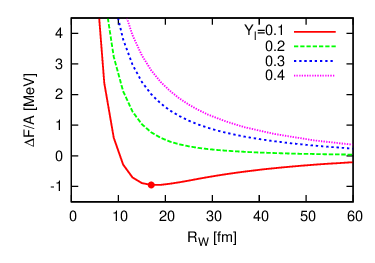}
\includegraphics[width=18pc]{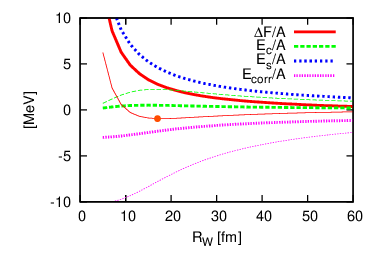}
\caption{\label{fig:03} Upper panel shows the stability curves for given lepton-number fractions as functions of the size of the Wigner-Seitz cell. The blob shows the minimum point of the free energy. Lower panel shows each contribution to the free energy. Thick lines show the case for $Y_l =0.4$, and thin lines for $Y_l =0.1$. 
}
\end{figure}

We can see how the pasta structures become mechanically unstable as $Y_l$ increases. 
The stability curves for slabs are shown for given $Y_l$ at $n_B=3 n_0$ in the upper panel of Fig.~\ref{fig:03}, where we show the free energy per baryon $\Delta F/A$ as a function of the cell radius $R_W$ by subtracting the asymptotic value in the case of $Y_l =0.1$. 
We find that the minimum point is shifted to a larger value as $Y_l$ increases, and eventually disappears beyond $Y_l \sim 0.2$; e.g., the nonuniform structure becomes mechanically unstable in the case of high $Y_l$, and the amorphous state develops.
Note that the mechanical instability is also brought about by the thermal effects, but it is mainly caused by neutrinos for PNS matter; 
actually we can see that the pasta structures still appear 
at the temperature under consideration, if neutrinos are absent \cite{yasutake09b}. 

This result can be understood as follows.
In the lower panel of Fig.~\ref{fig:03}, we show the contents of the free energy $\Delta F$ for $Y_l=0.1$ and $Y_l=0.4$.  
It consists of the Coulomb energy $E_{\rm C}/A$, the surface energy $E_{\rm S}/A$, and the correlation energy $E_{\rm corr}/A$ coming from 
 the change of the bulk energy due to the rearrangement of charge distribution \cite{voskresensky02}. 
Note that both $E_C$ and $E_{\rm corr}$ fully include the nonlinear effects of the Coulomb interaction.
The electron fraction becomes rich in the presence of neutrinos, which is accompanied by the enhancement of the fraction of positively charged particles as shown in Fig.\ref{fig:02}. 
The net charge density is then reduced over the whole region of the Wigner-Seitz cell, which leads to a large reduction of $E_C$ and $E_{\rm corr}$.
Consequently, the minimum point shifts to the larger $R_W$.

%

We have studied the HQ-phase transition and the properties of the mixed phase in PNSs by treating it as a ternary system. 
We have taken into account the finite-size effects properly, and calculated the density profiles in a self-consistent manner by imposing the Gibbs conditions on the phase equilibrium. The EOS for PNS matter becomes close to the one given by the bulk calculation (LCN) in the initial stage, even though we fully take into account the finite-size effects. This is because the crystalline structure of pasta is broken by the Coulomb interaction there to leave an amorphous state. Therefore, PNS matter is effectively reduced to a binary system; there appears no density 
discontinuity which is a specific feature in a pure system.

The pasta structures may appear in the middle stage of evolution of PNSs for $\sigma_S<$ 70 MeV~fm$^{-2}$, where initial cooling and deleptonization  
are well attained. 
PNS matter may be in nonequilibrium there. 
Hence, it needs numerical studies for the thermal and chemical evolution of PNSs including neutrino transport~\cite{fischer09} to know an exact era when the pasta structures appear. 
 For this study, the properties of PNS matter, e.g., amorphous state or pasta phases, are  important for the neutrino opacity.

 As a possible implication of our results one may consider the radiation of 
the gravitational wave from oscillating PNSs. It depends on the properties of the mixed phase: the shear modulus is brought about by the pasta structures~\cite{johnson11}  while the amorphous state is not.

Finally, we note again that the EOS has many uncertainties, especially for quark matter. 
We simply adopted the density-dependent bag model in this paper, 
while other quark models, including chiral restoration or color superconductivities, should be used for a realistic description of the phase transition~\cite{yasutake09a}.

\begin{acknowledgments}
We are grateful to H.-J.~Schulze, F.~G.~Burgio, and M.~Baldo for their warm hospitality and fruitful discussions. This work was partially supported by the Grant-in-Aid for the Global COE Program ``The Next Generation of Physics, Spun from Universality	and Emergence,'' from the Ministry of Education, Culture, Sports, Science and Technology (MEXT) of Japan, and the Grant-in-Aid for Scientific Research (C) (20540267, 19540313, 23540325). 
\end{acknowledgments}


\end{document}